\documentclass[prl,twocolumn,preprintnumbers,amsmath,amssymb]{revtex4-2}
\usepackage{graphicx}
\usepackage{dcolumn}
\usepackage{amsmath,amsbsy,amssymb,amsfonts,mathtools}
\usepackage{bm}
\usepackage[T1]{fontenc}
\usepackage{booktabs}
\usepackage{textcomp}
\usepackage{xcolor}
\usepackage{hyperref}
\usepackage{physics}
\hypersetup{
  pdfnewwindow=true,
  colorlinks=true,
  linkcolor=blue,
  anchorcolor=blue,
  citecolor=blue!80!green,
  filecolor=blue,
  menucolor=blue,
  urlcolor=blue
}

\begin{document}

\title{Real-space Floquet topology written by the orbital angular momentum of light}
%\title{Real-space Floquet topology imprinted by twisted light}

\author{Mohammad Shafiei}
\affiliation{COMMIT, Department of Physics \& NANOlight Center of Excellence,
University of Antwerp, Groenenborgerlaan 171, B-2020 Antwerp, Belgium}

\author{Milorad V.\ Milo\v{s}evi\'{c}}
\email{milorad.milosevic@uantwerpen.be}
\affiliation{COMMIT, Department of Physics \& NANOlight Center of Excellence,
University of Antwerp, Groenenborgerlaan 171, B-2020 Antwerp, Belgium}

\date{\today}

\begin{abstract}
Floquet engineering usually treats light as a uniform control field that changes the topology of an entire driven material. Here we show that structured light carrying orbital angular momentum (OAM) enables a different regime, in which topology is written directly in real space. For ultrathin topological insulator films, circularly polarized Laguerre--Gaussian beams generate a radial Floquet mass whose sign changes define a topological annulus bounded by two concentric chiral ring modes. The transition is helicity selective: below a thickness-dependent critical frequency, left-circularly polarized light drives mass inversion, whereas right-circularly polarized light increases the gap and leaves the film trivial. Independently, the OAM quantum number shifts and reshapes the annulus without changing the frequency, intensity, or helicity. In the decoupled-surfaces limit, the same mechanism produces a purely Floquet-induced topological mass and a vortex-core zero mode. These results identify photon OAM as a control parameter for nonequilibrium topology and provide a route to programmable topological landscapes in quantum materials.
\end{abstract}

\maketitle

\paragraph{Introduction.}
Floquet engineering has become a central route for creating quantum phases that have no equilibrium counterpart, by using periodic driving to renormalize band structures and induce effective gauge fields~\cite{rudner2020band,oka2009photovoltaic,kitagawa2011transport}. Circularly polarized light (CPL) is the canonical example: by breaking time-reversal symmetry it can transform Dirac materials into Floquet topological insulators (TIs), quantum anomalous Hall phases, and high-Chern-number states~\cite{shafiei2024floquet,rechtsman2013photonic,bao2022light}. More recently, linearly polarized light (LPL) was also shown to drive a Floquet topological transition in quasi-two-dimensional Dirac systems through the momentum-quadratic hybridization mass~\cite{shafiei2025linearly}. In all these settings, however, the optical drive is essentially a global knob: it changes the topology of the illuminated material as a whole~\cite{rudner2020band,shafiei2025light}.

Structured light suggests a qualitatively different possibility. Laguerre--Gaussian (LG) beams carry quantized orbital angular momentum $\ell\hbar$ per photon, encoded in a helical phase front and an annular intensity profile with a vortex singularity at the beam center~\cite{allen1992orbital,padgett2011tweezers,yang2022generation,bekshaev2011internal}. Optical vortices have already enabled synthetic gauge fields in atomic and photonic platforms~\cite{dalibard2011colloquium}. Whether the OAM of photons can instead be used to engineer spatially resolved Floquet topology in a quantum material remains largely unexplored.

Here we show that photon OAM promotes Floquet engineering from global band-structure control to the direct writing of topology in real space. We consider quasi-two-dimensional Dirac systems realized by ultrathin Bi$_2$Se$_3$ films~\cite{zhang2010crossover}. For such films, a circularly polarized Laguerre–Gaussian beam generates a radial Floquet mass whose sign reversals define a topological annulus bounded by two concentric chiral modes. Crucially, the optical degrees of freedom play distinct roles: helicity determines whether mass inversion is possible, through a thickness-dependent critical frequency, whereas the OAM quantum number $\ell$ independently controls the position and geometry of the topological region. Thus, topology can be created, displaced, and reshaped without changing the material or introducing static patterning. In the decoupled-surfaces limit, where the equilibrium mass vanishes, the optical vortex itself generates the topological mass and produces a vortex-core zero mode. Helicity therefore acts as a topological switch, while OAM determines where the switched phase resides in real space. In this way, structured light provides a route to optically writable and reconfigurable topological landscapes without static material patterning.

\paragraph{Model and Floquet mass.}
We start from the low-energy Hamiltonian~\cite{zhang2010crossover, zhang2012surface}
\begin{equation}
H_0(\mathbf{k})=
 v_F(k_y\sigma_x-k_x\sigma_y)
 +(\Delta_0-\Delta_1 k^2)\sigma_z,
\label{eq:H0}
\end{equation}
where $\sigma_i$ act in the surface-pseudospin subspace and $k^2=k_x^2+k_y^2$. The mass term $\Delta_0-\Delta_1 k^2$ describes intersurface hybridization in ultrathin TI films and may also include magnetic exchange contributions. The parameters used below correspond to experimentally characterized Bi$_2$Se$_3$ films~\cite{zhang2010crossover}, as summarized in Table~\ref{tab:params}. For thicknesses above approximately five quintuple layers (QLs), the two surfaces are effectively decoupled and we take $\Delta_0=\Delta_1=0$~\cite{zhang2010crossover, liu2010oscillatory}.

\begin{table}[b]
\centering
\caption{Material parameters for ultrathin Bi$_2$Se$_3$ films used in the calculations~\cite{zhang2010crossover}. The hybridization mass decreases with increasing thickness and vanishes in the decoupled-surfaces limit.}
\label{tab:params}
\renewcommand{\arraystretch}{1.12}
\begin{tabular}{c c c c c}
\toprule
Thickness &\,\, $\Delta_0$ (eV) &\,\, $\Delta_1$ (eV\AA$^2$) &\,\, $v_F$ (eV\AA) &\,\, regime \\
\midrule
2 QL &\,\, 0.126 &\,\, 21.8 &\,\, 3.10 &\,\, strong \\
3 QL &\,\, 0.069 &\,\, 18.0 &\,\, 3.17 &\,\, strong \\
4 QL &\,\, 0.035 &\,\, 10.0 &\,\, 2.95 &\,\, intermediate \\
5 QL &\,\, 0.020 &\,\, 5.0  &\,\, 2.99 &\,\, weak \\
$\geq6$ QL &\,\, 0.0 &\,\, 0.0 &\,\, 2.98 &\,\, decoupled \\
\bottomrule
\end{tabular}
\end{table}

The film is driven by a normally incident circularly polarized LG beam~\cite{allen1992orbital,yao2011orbital},
\begin{equation}
\begin{aligned}
\mathbf{A}(\mathbf{r},t)=&\,A_0
\left(\frac{\sqrt{2}r}{w}\right)^{|\ell|}e^{\frac{-r^2}{w^2}}
\left[\hat{\varepsilon}e^{i(\ell\varphi-\omega t)}
+\hat{\varepsilon}^{*}e^{-i(\ell\varphi-\omega t)}\right],
\end{aligned}
\label{eq:Avortex}
\end{equation}
where $w$ is the beam waist, $A_0$ the field amplitude, and $\hat{\varepsilon}=(\hat{x}\pm i\hat{y})/\sqrt{2}$ specifies the helicity. The plane-wave CPL limit is recovered for $\ell=0$. We introduce the drive through $\mathbf{k}\rightarrow \mathbf{k}+\mathbf{A}(\mathbf{r},t)$, and work in the off-resonant regime $\omega\gg\Delta_0$, appropriate for near-infrared photon energies $\hbar\omega=1.0$--$1.8$ eV.

Keeping the leading Floquet--Magnus term~\cite{eckardt2015high},
\begin{equation}
H_{\mathrm{eff}}=H_0+\frac{1}{\omega}[H_1,H_{-1}]+\mathcal{O}(\omega^{-2}),
\end{equation}
where $H_{\pm1}$ are the first Fourier harmonics of the driven Hamiltonian, yields
\begin{equation}
H_{\mathrm{eff}}=\tilde v_F(r)(k_y\sigma_x-k_x\sigma_y)
+[\tilde\Delta_0(r)-\Delta_1 k^2]\sigma_z .
\label{eq:Heff}
\end{equation}
All spatial dependence enters through the local vortex intensity
\begin{equation}
\mathcal{I}(r)=A_0^2
\left(\frac{\sqrt{2}r}{w}\right)^{2|\ell|}e^{-2r^2/w^2}.
\label{eq:intensity}
\end{equation}
For LCP driving one obtains
\begin{align}
\tilde v_F(r)&=v_F\left[1-\frac{2\Delta_1\mathcal{I}(r)}{\omega}\right],
\label{eq:vFr}\\
\tilde\Delta_0^{\mathrm{LCP}}(r)&=
\Delta_0-\mathcal{I}(r)\left(\Delta_1+\frac{v_F^2}{\omega}\right),
\label{eq:D0_lcp}
\end{align}
whereas RCP driving gives
\begin{equation}
\tilde\Delta_0^{\mathrm{RCP}}(r)=
\Delta_0-\mathcal{I}(r)\left(\Delta_1-\frac{v_F^2}{\omega}\right).
\label{eq:D0_rcp}
\end{equation}
Thus the helicity controls whether the Floquet commutator cooperates with or competes against the hybridization-induced mass renormalization. For comparison, LPL has vanishing first-order commutator and gives $\tilde\Delta_0^{\mathrm{LPL}}=\Delta_0-\Delta_1A_0^2/2$, recovering the transition discussed in Ref.~\cite{shafiei2025linearly}.

Equations~\eqref{eq:D0_lcp} and~\eqref{eq:D0_rcp} reveal a sharp helicity asymmetry. LCP always suppresses the mass gap. RCP instead changes character at
\begin{equation}
\omega^{*}=\frac{v_F^2}{\Delta_1}.
\label{eq:omstar}
\end{equation}
For $\omega<\omega^{*}$, RCP increases the mass and cannot invert it, while LCP can drive a topological transition. For $\omega>\omega^{*}$, both helicities reduce the gap, although LCP remains more efficient. The values of $\omega^{*}$ in Table~\ref{tab:omstar} show that 5 QL Bi$_2$Se$_3$ is especially favorable: $\hbar\omega^{*}=1.79$ eV lies in the visible range, so at $\hbar\omega=1.5$ eV the transition is fully helicity selective.

\begin{table}[b]
\centering
\caption{Helicity-selective crossover frequency $\omega^{*}=v_F^2/\Delta_1$. For $\omega<\omega^{*}$, RCP light increases the gap and cannot drive mass inversion, whereas LCP can.}
\label{tab:omstar}
\renewcommand{\arraystretch}{1.12}
\begin{tabular}{c c c c}
\toprule
Thickness &\,\, $\hbar\omega^{*}$ (eV) &\,\, $\lambda^{*}$ (nm) &\,\, range \\
\midrule
2 QL &\,\, 0.44 &\,\, 2813 &\,\, mid-IR \\
3 QL &\,\, 0.56 &\,\, 2221 &\,\, near-IR \\
4 QL &\,\, 0.87 &\,\, 1425 &\,\, near-IR \\
5 QL &\,\, 1.79 &\,\, 694 &\,\, visible \\
\bottomrule
\end{tabular}
\end{table}

For a spatially uniform LCP field, the transition occurs when $\tilde\Delta_0^{\mathrm{LCP}}=0$, yielding
\begin{equation}
A_{\mathrm{th}}^{\mathrm{LCP}}=
\sqrt{\frac{\Delta_0}{\Delta_1+v_F^2/\omega}}.
\label{eq:Ath_lcp}
\end{equation}
This threshold is lower than the LPL value $A_{\mathrm{th}}^{\mathrm{LPL}}=\sqrt{2\Delta_0/\Delta_1}$, showing that LCP provides both helicity selectivity and a more efficient route to Floquet mass inversion.

\begin{figure*}[t]
\centering
\includegraphics[width=0.98\textwidth]{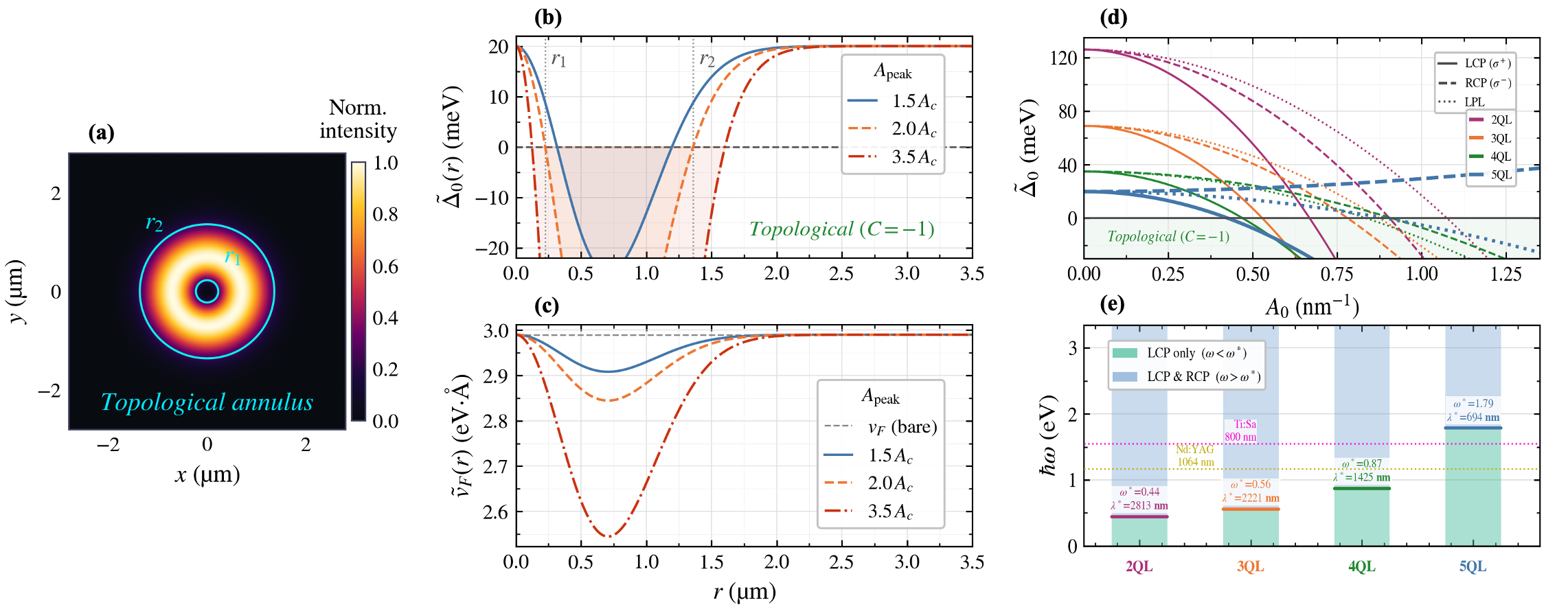}
\caption{Floquet mass engineering by circularly polarized Laguerre--Gaussian light. (a) Schematic of an $\ell=1$ optical vortex incident on a quasi-two-dimensional Bi$_2$Se$_3$ film. The annular intensity profile drives mass inversion only where the local field is large enough, producing a topological annulus bounded by the zero-mass contour $\tilde\Delta_0(r)=0$ (cyan). (b) Radial LCP mass profiles for a 5 QL film at $\hbar\omega=1.5$ eV and $w=1~\mu$m, showing the two domain walls $r_1$ and $r_2$. (c) Corresponding spatial renormalization of the Fermi velocity. (d) Uniform-drive mass renormalization for LCP, RCP, and LPL fields. For a 5 QL film, LCP closes the gap while RCP increases it at $\hbar\omega=1.5$ eV. (e) Thickness dependence of $\omega^{*}=v_F^2/\Delta_1$, identifying the 5 QL film as the visible-frequency helicity-selective regime.}
\label{fig:fig1}
\end{figure*}

\paragraph{Floquet topological annulus.}
The LG intensity in Eq.~\eqref{eq:intensity} vanishes at the optical vortex core, peaks at $r_{\max}=w\sqrt{|\ell|/2}$, and decays at large radius. Consequently $\tilde\Delta_0(0)=\Delta_0>0$: the core remains trivial. If LCP illumination is strong enough that $\tilde\Delta_0(r_{\max})<0$, the mass changes sign twice, at an inner radius $r_1$ and an outer radius $r_2$. The region $r_1<r<r_2$ is then a Floquet topological annulus with Chern number $C=-1$, while both the core and the exterior remain trivial. Bulk-boundary correspondence requires a pair of one-dimensional chiral modes localized at the two circular domain walls. Since the two interfaces have opposite orientations, the inner and outer ring modes counter-propagate.

Figures~\ref{fig:fig1}(a)--(c) show the real-space Floquet response for a 5~QL Bi$_2$Se$_3$ film driven by an LCP Laguerre--Gaussian beam with $\ell=1$, $w=1~\mu$m, and $\hbar\omega=1.5$~eV. Panel (a) shows the annular optical profile and the zero-mass contour that defines the topological region. Panels (b) and (c) show how increasing the peak drive amplitude deepens the negative-mass region and suppresses $\tilde v_F(r)$ near the optical maximum, while both quantities return to their undriven values near the core and far outside the beam. The global Floquet phase behavior under spatially uniform illumination is summarized in Figs.~\ref{fig:fig1}(d) and \ref{fig:fig1}(e). Figure~\ref{fig:fig1}(d) presents the Floquet-renormalized Dirac mass $\tilde{\Delta}_0$ as a function of the driving amplitude $A_0$ for 2--5~QL Bi$_2$Se$_3$ films under LCP, RCP, and linearly polarized illumination at $\hbar\omega=1.5$~eV. The critical amplitudes marking the onset of the Floquet topological transition are indicated for each polarization and film thickness. Figure~\ref{fig:fig1}(e) shows the thickness dependence of the helicity-selective crossover frequency $\omega^{*}$, identifying the 5~QL film as the unique visible-frequency regime in which the Floquet transition becomes fully helicity selective.

The annulus nucleates when the mass first vanishes at $r_{\max}$. Since $\mathcal{I}(r_{\max})=A_0^2 |\ell|^{|\ell|}e^{-|\ell|}$, the critical amplitude is
\begin{equation}
A_c(\ell)=
\sqrt{\frac{\Delta_0}{|\ell|^{|\ell|}e^{-|\ell|}
(\Delta_1+v_F^2/\omega)}} .
\label{eq:Ac}
\end{equation}
For $\ell=1$, this reduces to $A_c=\sqrt{e\Delta_0/(\Delta_1+v_F^2/\omega)}$, where $e$ is the base of the natural logarithm. Above the threshold, the domain-wall radii are determined by
\begin{equation}
\mathcal{I}(r_{1,2})
\left(\Delta_1+\frac{v_F^2}{\omega}\right)=\Delta_0,
\label{eq:rstar}
\end{equation}
and are obtained numerically.

\begin{figure}[b]
\centering
\includegraphics[width=0.90\linewidth]{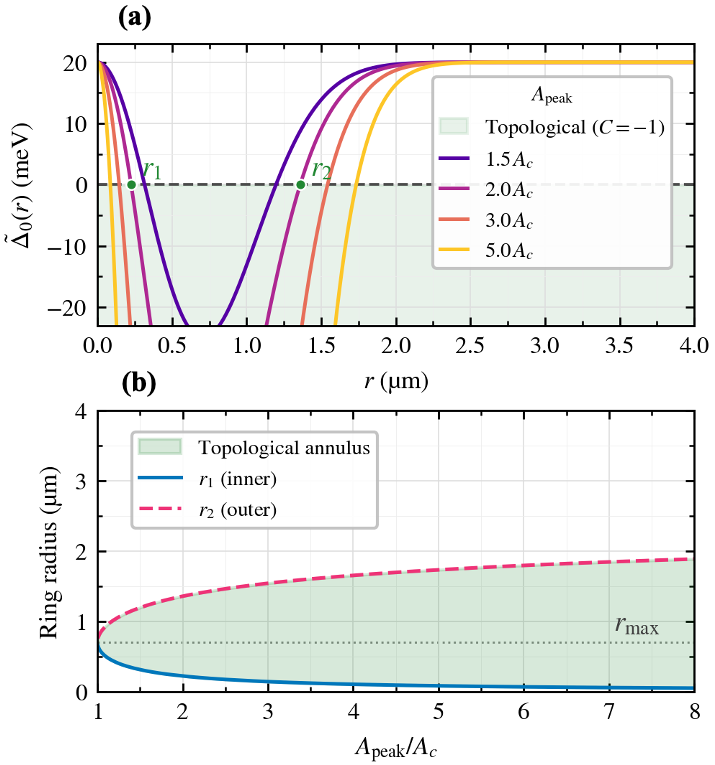}
\caption{Formation and expansion of the Floquet topological annulus. (a) Radial mass profiles for a 5 QL film driven by an LCP LG beam with $\ell=1$, $w=1~\mu$m, and $\hbar\omega=1.5$ eV. The shaded region denotes the inverted-mass annulus. (b) Inner and outer domain-wall radii as functions of $A_{\mathrm{peak}}/A_c$; the dotted line marks the optical-intensity maximum $r_{\max}=w/\sqrt{2}$.}
\label{fig:fig2}
\end{figure}

Figure~\ref{fig:fig2} quantifies the evolution of the annulus in a 5 QL film. Just above the threshold, the two radii coincide near $r_{\max}$. Increasing the drive pushes $r_1$ inward and $r_2$ outward, thereby widening the topological region. For $A_0=2A_c$, we find $r_1=0.23~\mu$m and $r_2=1.36~\mu$m, corresponding to circumferences of $1.4~\mu$m and $8.5~\mu$m. These length scales are compatible with spatially resolved near-field and scanning-probe measurements.

Restoring SI units, the peak intensity at the threshold is
\begin{equation}
I_c=\frac{1}{2}nc\varepsilon_0
\left(\frac{\hbar\omega A_c}{q_e}\right)^2,
\end{equation}
where $q_e$ is now the elementary charge, $n\simeq1$ is the refractive index for near-normal incidence from vacuum, and $c$ denotes the speed of light in vacuum, and $\varepsilon_0$ is the vacuum permittivity. For a 5 QL film at $\hbar\omega=1.5$ eV and $w=1~\mu$m, this gives $I_c\simeq0.15~\mathrm{TW/cm^2}$, within the range used in ultrafast Floquet experiments on TI surface states~\cite{gedik2017photoemission,wang2013observation}.

\paragraph{Control by photon OAM.}
Photon OAM provides an independent geometric control knob. Increasing $|\ell|$ shifts the optical maximum as $r_{\max}=w\sqrt{|\ell|/2}$ and therefore moves the two topological interfaces away from the beam center without changing the frequency, polarization, or peak-intensity scale of the Floquet mechanism.

\begin{figure}[t]
\centering
\includegraphics[width=0.49\textwidth]{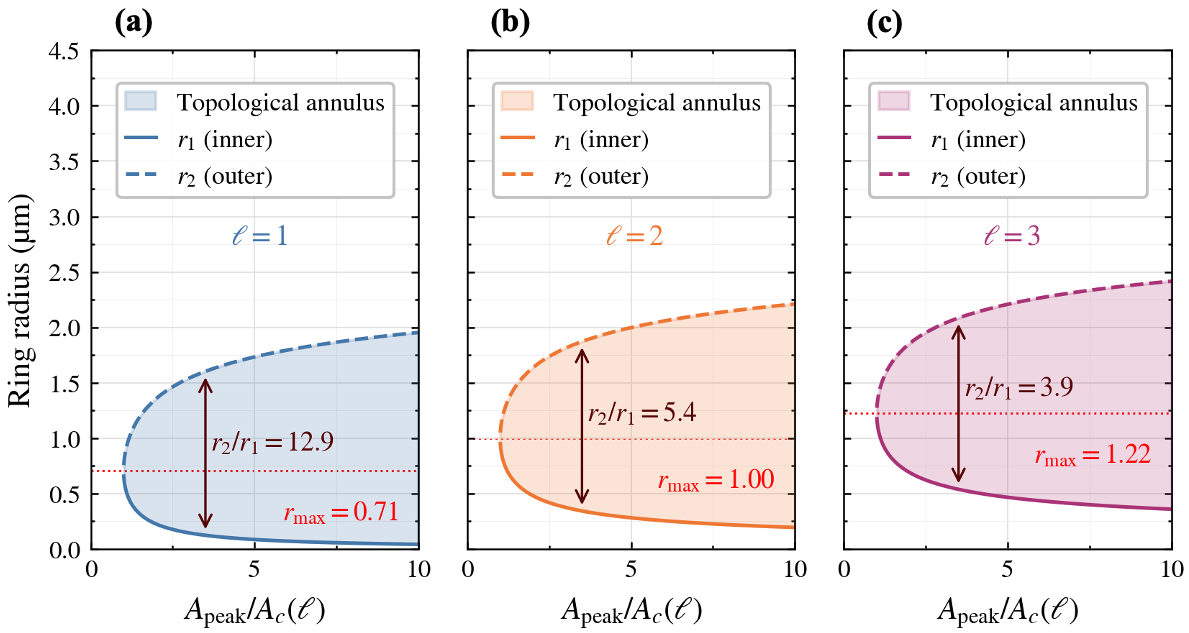}
\caption{OAM control of the topological annulus. Inner ($r_1$, solid) and outer ($r_2$, dashed) domain-wall radii are shown versus $A_{\mathrm{peak}}/A_c(\ell)$ for $\ell=1,2,3$ in a 5 QL film at $\hbar\omega=1.5$ eV and $w=1~\mu$m. The dotted curves mark $r_{\max}=w\sqrt{\ell/2}$. Larger OAM shifts the annulus outward and changes its relative width, as indicated by the double arrows evaluated at $A_{\mathrm{peak}}/A_c=3.5$.}
\label{fig:fig3}
\end{figure}

This behavior is shown in Fig.~\ref{fig:fig3}. After normalizing each curve to its own $A_c(\ell)$, the characteristic radius grows approximately as $\sqrt{|\ell|}$. At fixed $A_{\mathrm{peak}}/A_c(\ell)$, both $r_1$ and $r_2$ move outward as $\ell$ increases. The relative width is also reshaped: at $A_{\mathrm{peak}}/A_c=3.5$, the ratio $r_2/r_1$ decreases from approximately $12.9$ for $\ell=1$ to $5.4$ for $\ell=2$ and $3.9$ for $\ell=3$. Thus OAM does not merely translate the topological region; it also modifies its geometry.

\begin{figure}[t]
\centering
\includegraphics[width=0.49\textwidth]{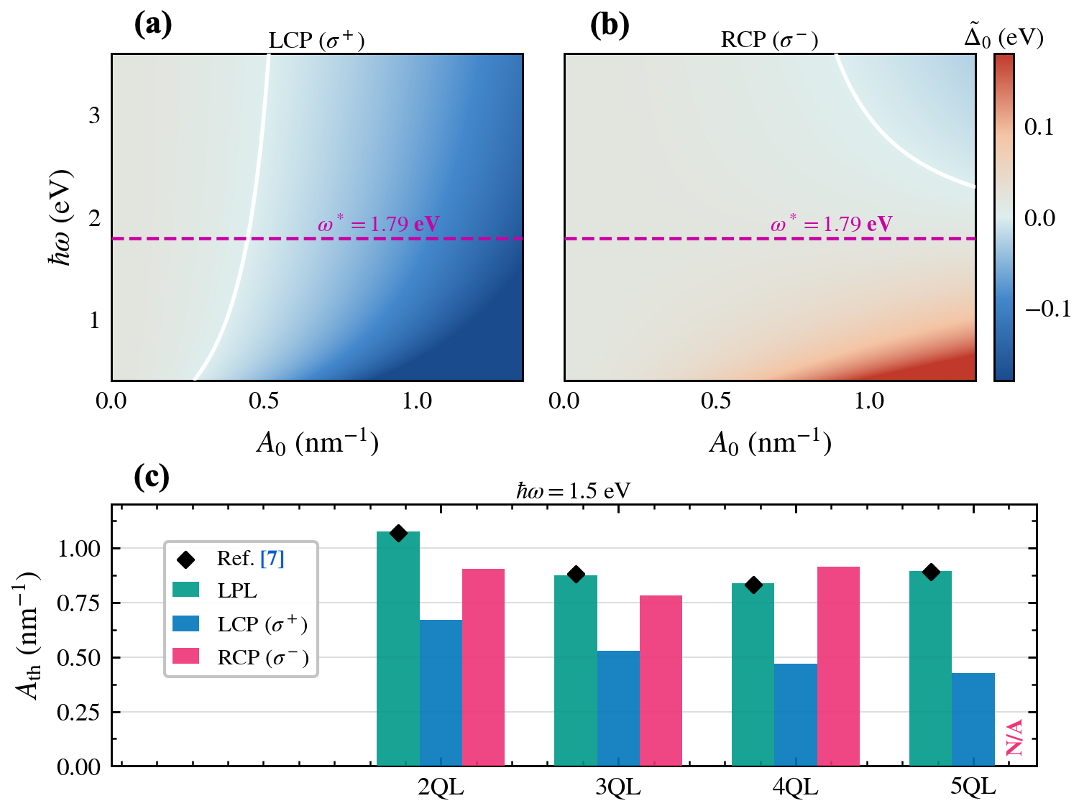}
\caption{Helicity-selective Floquet phase diagrams. (a) LCP phase diagram of a 5 QL film in the $(A_0,\hbar\omega)$ plane. Blue and red indicate topological ($C=-1$) and trivial ($C=0$) regions, respectively; the white contour is $\tilde\Delta_0=0$, and the dashed line marks $\omega^{*}$. (b) RCP phase diagram. No topological phase occurs below $\hbar\omega^{*}=1.79$ eV. (c) Critical amplitudes for LPL~\cite{shafiei2025linearly}, LCP, and RCP illumination at $\hbar\omega=1.5$ eV. For 5 QL, RCP has no transition at this frequency.}
\label{fig:fig4}
\end{figure}

The decoupled-surfaces limit, corresponding to $\Delta_0=\Delta_1=0$, is qualitatively distinct. In this case the equilibrium mass vanishes, and the gap is generated entirely by the Floquet commutator:
\begin{equation}
\tilde\Delta_0^{\geq6}(r)=
-\frac{v_F^2}{\omega}\mathcal{I}(r)\le0.
\label{eq:6QL}
\end{equation}
The optical vortex therefore creates a purely light-induced topological mass everywhere except at the singular core, where the intensity vanishes. The core acts as the only topological boundary and binds a symmetry-protected zero mode, analogous to a Caroli--de~Gennes--Matricon vortex state~\cite{caroli1964bound}, but generated here by photon OAM rather than by superconducting phase winding.

\paragraph{Observable signatures.}
The most direct signature is the helicity-controlled appearance of two concentric in-gap ring states. Under LCP illumination, the topological annulus forms between $r_1$ and $r_2$; below $\omega^{*}$, switching to RCP removes the annulus and restores a trivial insulating gap. Figure~\ref{fig:fig4} summarizes this behavior through LCP and RCP phase diagrams for a 5 QL film and compares the corresponding transition thresholds with the LPL case.

Scanning tunneling spectroscopy or near-field optical spectroscopy should detect enhanced local density of states at the two domain walls, with a separation on the micron scale~\cite{wang2013observation} for the parameters shown in Fig.~\ref{fig:fig2}. Transport devices patterned with contacts intersecting the annulus would show helicity-dependent conducting channels: LCP activates chiral ring transport, whereas RCP suppresses it in the helicity-selective regime. The two counter-propagating circular modes should also generate spatially structured photocurrents and magnetic responses, observable through time-resolved terahertz emission or magneto-optical Kerr microscopy~\cite{kirilyuk2010ultrafast}. These signatures require no magnetic field or electrostatic reconfiguration; the switch is implemented solely by the helicity and OAM of the pump.

\paragraph{Conclusion.}
We showed that structured light can do more than renormalize a Floquet band structure: it can program topology directly in real space, as demonstrated here by the helicity-selective formation and OAM-controlled geometry of a topological annulus. Since helicity controls the onset of mass inversion while photon OAM sets the position and shape of the resulting interfaces, topological regions can be written, displaced, reshaped, or erased without static material patterning. The demonstrated OAM control of the two domain walls suggests that superpositions of optical vortices or more general structured wavefronts could extend the annular geometry to tailored topological domains and networks of chiral channels, while time-dependent beam shaping may enable controlled motion and coupling of Floquet boundary states. In the decoupled-surfaces regime, the vortex-core zero mode generated by the purely Floquet-induced mass similarly points toward arrays of optical vortices as a route to create and couple localized states. Establishing the robustness of these optically defined structures against disorder, finite pulse duration, and heating, and extending the mechanism to other gapped Dirac materials, could generalize photon OAM as a control variable for programmable nonequilibrium topology and topological circuitry.

\paragraph{Acknowledgments.}
This work was supported by the Research Foundation--Flanders (FWO-Vlaanderen), the FWO-FNRS EOS project ShapeME, and the Special Research Funds (BOF) of the University of Antwerp.

\bibliography{bibliography}

\end{document}